# The Impact of Farmers' Borrowing Behavior on Agricultural Production Technical Efficiency

Hambur Wang

Guanghua School of Management, Peking University, Beijing 100871, China

**Abstract:** The effectiveness of farmer loan policies is crucial for the high-quality development of agriculture and the orderly advancement of the rural revitalization strategy. Exploring the impact of farmers' borrowing behavior on agricultural production technical efficiency holds significant practical value. This paper utilizes data from the 2020 China Family Panel Studies (CFPS) and applies Stochastic Frontier Analysis (SFA) along with the Tobit model for empirical analysis. The study finds that farmers' borrowing behavior positively influences agricultural production technical efficiency, with this effect being especially pronounced among low-income farmers. Additionally, the paper further examines household characteristics, such as household head age, gender, educational level, and the proportion of women in the family, in relation to agricultural production technical efficiency. The findings provide policy recommendations for optimizing rural financial service systems and enhancing agricultural production technical efficiency.

## 1. Introduction

Amid the backdrop of agricultural modernization and rural revitalization, the Chinese government places a high priority on food security and the high-quality development of agriculture. It emphasizes the implementation of the rural revitalization strategy, the refinement of institutional policies for prioritizing agricultural and rural development, and actively promotes integrated urban-rural development. Improving agricultural production technical efficiency is essential for ensuring national food security. Achieving this goal depends not only on the capabilities of agricultural producers themselves but also on the support of the rural financial service system. In recent years, China has established a multi-layered rural financial system, transitioning from traditional to emerging forms, and the implementation of farmer loan policies has become a crucial condition for promoting industrial development. However, there is limited research on the specific impact of farmers' borrowing behavior on agricultural production technical efficiency. As financial support continues to expand, investigating how farmers' borrowing behavior affects agricultural production technical efficiency holds significant value for improving the supply and service system of rural finance in China, as well as for ensuring stable food supply and promoting high-quality agricultural development.

**Existing Research on Farmers' Borrowing and Agricultural Production Technical Efficiency:**

The current literature provides a rich theoretical and empirical foundation on both farmers' borrowing and agricultural production technical efficiency. Studies on farmers' borrowing behavior primarily focus on influencing factors, income-enhancing effects, and impacts on input of production factors. In terms of influencing factors, some studies suggest that farmers' borrowing is influenced by individual endowments, household characteristics, social capital, and regional financial development [1,2,3]. Numerous studies demonstrate that loans from formal financial institutions positively impact household income [4,5] and that borrowing by farmers mainly affects the input of physical capital and land scale [6,7]. Concerning agricultural production technical efficiency, the literature primarily addresses measurement methods, influencing factors, and spatiotemporal changes, with commonly used measurement methods including Data Envelopment Analysis (DEA) and Stochastic Frontier Analysis (SFA) [8,9].

**International and Domestic Perspectives on Financial Markets and Production Efficiency:**

Internationally, some scholars have examined the relationship between financial markets and production efficiency, finding that access to financing is positively associated with production efficiency and that credit availability has a significant positive effect on total factor productivity [10,11]. Additionally, the impact of agricultural credit on farmers' technical efficiency is complex and varies significantly based on the specific circumstances of farmers [12,13,14]. In contrast, domestic scholars tend to focus more on efficiency studies, finding that China's rural financial resource allocation efficiency is low and that there are significant regional disparities [15,16]. Research also indicates that the formal financial system suppresses rural financial demand to a certain extent, implying that rural financial efficiency requires improvement [17].

While existing literature provides theoretical and methodological guidance for analyzing the impact of farmers' borrowing on agricultural production technical efficiency, most studies examine the macro perspective of rural financial development's role in enhancing technical efficiency. Few studies directly link farmers' borrowing with agricultural production technical efficiency on a micro level. This paper takes a micro-level approach by analyzing CFPS data to examine the impact of farmers' borrowing on agricultural production technical efficiency and explores the heterogeneous effects of different types of borrowing on technical efficiency. This study aims to offer new insights into understanding and improving the relationship between farmers' borrowing behavior and production efficiency.

## 2. Theoretical Analysis

In China's agricultural production system, households, as the basic units contracting land, utilize funds to acquire various production factors to achieve agricultural output. This process can be viewed as a production entity operating within a perfectly competitive market. Based on neoclassical economic theory, each farmer, as a rational producer, aims to maximize profits. Farmers input various production factors, facing different factor prices, and strive to reach the maximum possible output frontier. Within this framework, technical efficiency (TE) refers to finding the optimal combination of input factors to bring output as close as possible to the production frontier. Any output system that fails to reach the output frontier can be considered technically inefficient (TI).

China's agricultural production system is typically more output-oriented, meaning that after determining the quantity of agricultural inputs, farmers seek to optimize output with fixed input levels. Farmers acquire production inputs for agricultural production through financial credit; however, loans are not always used exclusively for production. They may also cover debt repayment, rural customary expenses, housing costs, and other expenditures that are essential in rural society. Consequently, the impact of credit input on agricultural production may be either positive or negative. Additionally, even when credit is directed toward agricultural production, the allocation of funds—such as for purchasing productive materials, intermediate inputs, or hiring labor—can lead to different impacts on production efficiency [18].

Overall, assuming that farmers are rational agents, in the short term, borrowed funds can address temporary household liquidity constraints. Even without an outward shift in the production frontier, farmers can alleviate their financial constraints through borrowing, which may enhance agricultural production technical efficiency to some extent. In the long term, as the production frontier continuously shifts outward with new technological advancements, farmers can achieve higher production technical efficiency through rational production decision-making processes [9].

## 3. Data, Variables, and Model

### 3.1 Data Source and Variable Selection

The data used in this study are sourced from the 2020 China Family Panel Studies (CFPS) database, which surveys all members of sample households, covering 804 counties and 2,588 townships across 31 provinces in China. After filtering out urban households and missing data, the effective sample size is 5,541. The variable selection is presented in Table 1.

**Table 1: Variable Selection, Definitions, and Values**

| Variable Type | Variable Name | Definition and Values | Mean | Standard Deviation |
| --- | --- | --- | --- | --- |
| Input Indicators | Labor Input | Agricultural labor participation by family members | 0.71 | 0.48 |
| | Farm Size | Actual cultivated area, considering collective allocation, contracting, and transferred land area (mu) | 2.10 | 1.07 |
| | Input of Production Materials | Costs of seeds, fertilizers, pesticides, irrigation, and other expenses (CNY) | 7.80 | 1.42 |
| | Machinery Input | Total value of agricultural machinery and rental costs (CNY) | 7.82 | 1.51 |
| Output Indicator | Total Household Agricultural Output | Total market value of household agricultural products and self-consumption value (CNY) | 9.05 | 1.42 |
| Dependent Variable | Household Production Technical Efficiency | Measured using Stochastic Frontier Analysis (SFA) | 0.49 | 0.15 |
| Independent Variable | Household Borrowing Behavior | 0 = No, 1 = Yes | 0.16 | 0.36 |
| Control Variables | Household Head Age | 2020 minus birth year (years) | 52.33 | 19.45 |
| | Household Head Gender | Male = 1; Female = 0 | 0.51 | 0.50 |
| | Household Head Education | Years of formal education received by household head (years) | 5.44 | 6.33 |
| | Average Age | Total age of members divided by household size (years) | 46.64 | 17.03 |
| | Proportion of Female Family Members | Female members as a proportion of household members | 0.47 | 0.29 |
| | Family Financial Assets | Total value of household financial assets (logarithmic form, CNY) | 9.81 | 3.78 |
| | Topography | 1 = Hill; 2 = Mountain; 3 = Plain; 4 = Other | 2.53 | 6.47 |

## 3.2 Model Specification

This study employs the Stochastic Frontier Analysis (SFA) to estimate the technical efficiency of household agricultural production. SFA considers both controllable and uncontrollable stochastic factors in agricultural production. The error term includes both a technical inefficiency term and a random error term, which aligns better with the characteristics of agricultural production. Technical efficiency reflects the efficiency of input utilization and allocation in agricultural production. The basic theoretical model is:

$$\ln q_i = \beta_0 + \beta_1 \ln x_i + v_i - u_i$$

where $q_i$ represents the actual output of sample iii, and $x_i$ represents the input factors for sample iii. The term $v_i$ represents uncontrollable factors in production, capturing measurement error and random disturbances, where $v_i \sim N(0, \sigma^2)$; $u_i$ represents the inefficiency component, indicating the distance between the sample output and the production frontier, following a half-normal distribution. Based on this, the technical efficiency expression is:

$$TE_i = \frac{q_i}{\exp(x_i'\beta + v_i)} = \frac{\exp(x_i'\beta + v_i - u_i)}{\exp(x_i'\beta + v_i)} = \exp(-u_i)$$

where the technical efficiency $TE_i$ is calculated by isolating the inefficiency $u_i$ from the composite error term $v_i - u_i$.

To study the impact of household borrowing on agricultural production technical efficiency, the following baseline regression model is established based on the theoretical analysis presented above:

$$TE_i = \beta_0 + \beta_1 \text{ Loan }_i + \beta_2 X_i + \mu_i$$

where $TE_i$ is the agricultural production technical efficiency of household iii, $Loan_i$ indicates whether household i has engaged in borrowing behavior, and $X_i$ represents control variables.

## 4. Empirical Results and Analysis

### 4.1 Technical Efficiency of Farmers' Agricultural Production

Table 2: Technical Efficiency Estimation Results Using SFA Method

| Variable | Coefficient | Standard Error |
|---|---|---|
| Agricultural Output Value (log) | - | - |
| Input of Production Materials (log) | 0.622*** | (0.013) |
| Farm Size (log) | 0.068*** | (0.013) |
| Agricultural Machinery Input (log) | 0.169*** | (0.010) |
| Labor Input (log) | 0.087*** | (0.018) |
| Constant | 3.383*** | (0.100) |

**Sample Size:** 5541     **LR Test Statistic:** 450

Using the Stochastic Frontier Analysis (SFA) method, we measured the technical efficiency of farmers' agricultural production, with results presented in Tables 2 and 3. The LR test statistic of 450 significantly exceeds the critical value, indicating the presence of technical inefficiency and validating the model's effectiveness in assessing farmers' production efficiency. Inputs in agricultural machinery, labor, farm size, and production materials all exhibit positive coefficients at a 1% significance level, suggesting that increases in these inputs lead to improved technical efficiency in production, all else being equal.

Table 3: Differences in Technical Efficiency by Farmers' Loan Status

| Technical Efficiency Range | Number of Loan Borrowers | Percentage (%) | Number of Nonborrowers | Percentage (%) |
|---|---|---|---|---|
| TE<0.5 | 924 | 44.9 | 1760 | 50.6 |
| 0.5<=TE<0.6 | 618 | 30.0 | 1203 | 34.6 |
| 0.6<=TE<0.7 | 382 | 18.5 | 382 | 11.0 |
| 0.7<=TE<0.8 | 108 | 5.2 | 108 | 3.1 |
| 0.8<=TE<0.9 | 28 | 1.3 | 28 | 0.8 |
| 0.9<=TE<1.0 | 0 | 0 | 0 | 0 |
| Maximum Value | 0.873 | — | 0.866 | — |
| Minimum Value | 0.333 | — | 0.201 | — |
| Mean | 0.590 | — | 0.502 | — |
| Standard Deviation | 0.12 | — | 0.15 | — |
| Sample Size | 2060 | — | 3481 | — |

The average technical efficiency of sampled farmers in agricultural production is 0.532, indicating a degree of efficiency loss. From the perspective of borrowing, the technical efficiency of loan borrowers is 0.590, compared to 0.502 for non-borrowers, suggesting that borrowing behavior positively contributes to agricultural technical efficiency to some extent. While the maximum efficiency levels are similar between groups, the minimum efficiency values differ considerably, indicating that borrowing may significantly enhance the technical efficiency of less efficient farmers.

## 4.2 Impact of Farmers' Borrowing on Agricultural Production Technical Efficiency

Table 4: Regression Estimates of the Impact of Farmers' Borrowing on Technical Efficiency

| Variable | (1) | (2) | (3) |
|---|---|---|---|
| Loan Status | 0.013$^{**}$ | 0.021$^{***}$ | 0.024$^{***}$ |
|  | (0.004) | (0.005) | (0.006) |
| Household Head's Age | — | -0.000 | -0.000 |
|  | — | (0.000) | (0.000) |
| Household Head's Gender | — | -0.005 | -0.002 |
|  | — | (0.004) | (0.005) |
| Household Head's Education Level | — | -0.001 | -0.000 |
| | — | (0.001) | (0.001) |
| Average Age of Household | — | -0.001*** | -0.001*** |
| | — | (0.000) | (0.000) |
| Female Proportion in Household | — | 0.031** | 0.056*** |
| | — | (0.009) | (0.013) |
| Household Financial Assets | — | 0.000** | 0.000** |
| | — | (0.000) | (0.000) |
| Village Topography | — | — | 0.001 |
| | — | — | (0.000) |
| Constant | 0.498*** | 0.523*** | 0.518*** |
| | (0.002) | (0.010) | (0.014) |
| Sample Size | 5541 | 5541 | 5541 |

The results from the regression analysis in Table 4 show that farmers' loan status is positively and statistically significant across all three model specifications, indicating that borrowing has a positive effect on technical efficiency. Borrowing farmers perform closer to the production frontier than non-borrowing farmers, underscoring the potential positive impact of financial services on improving agricultural production efficiency.

Notably, both the average age of the household and female proportion are significant in the models. Specifically, the household average age coefficient is negative and significant at the 1% level, indicating that increased average age within households correlates with reduced technical efficiency, likely reflecting lower productivity or slower adoption of new technology. Meanwhile, a higher female proportion in the household positively correlates with technical efficiency, suggesting that women contribute positively to agricultural productivity, likely through effective management, resource allocation, and day-to-day agricultural activities.

## 4.3 Heterogeneity Analysis

Table 5: Heterogeneity Analysis by Household Total Income Level

| Variable | (1) Below Average Income | (2) Above Average Income |
|---|---|---|
| Loan Status | 0.022** | 0.027** |
| | (0.008) | (0.010) |
| Household Head's Age | 0.000 | -0.000 |
| | (0.000) | (0.000) |
| Household Head's Gender | -0.003 | -0.000 |
| | (0.007) | (0.009) |
| Household Head's | 0.000 | -0.001 |

|  | (0.001) | (0.001) |
| --- | --- | --- |
| Education Level | | |
| Average Age of Household | -0.001*** | -0.001* |
|  | (0.000) | (0.000) |
| Female Proportion in Household | 0.067*** | 0.031 |
|  | (0.016) | (0.021) |
| Household Financial Assets | 0.000* | 0.000 |
|  | (0.000) | (0.000) |
| Village Topography | 0.001** | -0.001 |
|  | (0.001) | (0.001) |
| Constant | 0.504*** | 0.546*** |
|  | (0.017) | (0.023) |
| Sample Size | 3288 | 2253 |

The analysis by household income reveals heterogeneity in the impact of borrowing on technical efficiency. Among households below average income, borrowing significantly improves technical efficiency, indicating that loans have a critical role in enhancing technical efficiency when resources are limited. For higher-income households, loans also improve technical efficiency but to a lesser extent, likely due to greater self-financing capability.

**Table 6: Heterogeneity Analysis by Farm Size**

|  | (1) Above Average Farm Size | (2) Below Average Farm Size |
| --- | --- | --- |
| Loan Status | 0.031*** | 0.013 |
|  | (0.008) | (0.009) |
| Household Head's Age | -0.000 | 0.000 |
|  | (0.000) | (0.000) |
| Household Head's Gender | -0.001 | -0.004 |
|  | (0.007) | (0.008) |
| Household Head's Education Level | -0.001 | 0.001 |
|  | (0.001) | (0.001) |
| Average Age of Household | -0.001*** | -0.000 |
|  | (0.000) | (0.000) |
| Female Proportion in Household | 0.054*** | 0.057* |
|  | (0.015) | (0.022) |
| Household Financial Assets | 0.000* | 0.000 |
|  | (0.000) | (0.000) |
| Village Topography | 0.002* | 0.000 |
|  | (0.001) | (0.001) |
| Constant | 0.545*** | 0.473*** |
|  | (0.017) | (0.023) |

|  | Sample Size | 3082 | 2459 |
|---|---|---|---|

Farm size heterogeneity analysis shows that borrowing has a stronger positive effect on technical efficiency for larger farms, suggesting that these farms may use loans more effectively for technological advancement.

## 4.4 Robustness Check

**Table 7: Robustness Check Using Alternative Variables**

|  | (1) | (2) |
|---|---|---|
| Loan Status | 0.024*** | — |
|  | (0.006) | — |
| Household Head's Age | — | 0.055*** |
|  | — | (0.011) |
| Household Head's Gender | -0.000 | -0.000 |
|  | (0.000) | (0.001) |
| Household Head's Education Level | -0.002 | -0.011 |
|  | (0.005) | (0.013) |
| Average Age of Household | -0.000 | -0.001 |
|  | (0.001) | (0.002) |
| Female Proportion in Household | -0.001*** | -0.002** |
|  | (0.000) | (0.001) |
| Household Financial Assets | 0.056*** | 0.092* |
|  | (0.013) | (0.039) |
| Household Financial Assets | 0.000** | -0.000 |
|  | (0.000) | (0.000) |
| Household Non-mortgage Financial Debt | 0.000* | 0.000*** |
|  | (0.000) | (0.000) |
| Village Topography | 0.001 | 0.001 |
|  | (0.000) | (0.001) |
| Constant | 0.518*** | 1.097*** |
|  | (0.014) | (0.118) |
| $N$ | 5541 | 5541 |

To validate the robustness of our findings, we replaced the binary variable of loan status with the continuous variable of loan amount and applied a Tobit model. The results (Table 7) show that loan amount negatively impacts technical efficiency at the 1% significance level, consistent with our earlier conclusions, reinforcing the robustness of this study's findings.

## 5. Conclusion and Policy Recommendations

This study analyzed the impact of household borrowing behavior on agricultural production technical efficiency. The findings indicate that the technical efficiency of borrowing households is higher than that of non-borrowing households, highlighting the positive role of loans in enhancing technical efficiency among less efficient households. Regression results show that loans significantly and positively promote technical efficiency, especially among low-income households and those operating at larger scales. The age of the household head negatively affects technical efficiency, while a higher proportion of women in the household correlates positively with efficiency, underscoring the positive contributions of women in agricultural production. Additionally, the value of household financial assets and the topography of the village significantly impact technical efficiency in larger households. Overall, loans are a crucial factor in promoting agricultural technical efficiency, particularly for low-income and large-scale households, as they provide essential financial support to improve production efficiency. Consequently, financial services play a critical role in enhancing agricultural production efficiency, and appropriate financial products and services can help improve technical efficiency and the production frontier for farming households.

Based on these findings, this paper proposes the following policy recommendations: First, tailored rural financial services should be designed to meet diverse borrowing needs. This includes increasing policy-based financial institution support for agricultural production, expanding the investment from commercial banks and other financial institutions in the agricultural sector, and utilizing the role of small- and medium-sized financial institutions, such as rural credit cooperatives and village banks, in supporting agriculture. Additionally, innovation in rural financial service products and collateralized loans for farmers should be encouraged to mitigate moral hazard and establish a precise rural financial information management system to address farmers' challenges of "difficult and expensive financing." Second, agricultural technical training should be conducted to optimize the efficiency of loan fund allocation. Through training, farmers' mastery of new technologies and knowledge will improve, enhancing their ability to optimize production factor allocation and thereby improving loan fund allocation efficiency. Governments and financial institutions should guide farmers in the appropriate use of borrowed funds, promote high-quality, high-yield crop varieties, and provide guidance on pesticide and fertilizer selection and use to reduce environmental pollution and soil degradation risks. Lastly, rural labor should be encouraged to seek non-agricultural employment to ease household production capital constraints. By promoting non-agricultural employment information and providing technical training, rural laborers can enhance their human capital and increase non-agricultural employment opportunities. Within the context of the Rural Revitalization Strategy, support for non-agricultural industry development should be provided to create local non-agricultural employment opportunities, shift surplus rural labor, address land-labor conflicts, promote agricultural production scale and mechanization, and enhance technical efficiency and industry competitiveness in agriculture. In future research, deep learning algorithms such as transformers and Mamba [19, 20, 21] can be incorporated into prediction models, along with attention mechanisms [22] to improve model performance and prediction accuracy.